\pdfoutput=1
\documentclass{article}

\usepackage[preprint]{neurips_2026}

\usepackage[T1]{fontenc}
\usepackage[utf8]{inputenc}
\usepackage{hyperref}
\usepackage{url}
\usepackage{booktabs}
\usepackage{amsfonts}
\usepackage{amsmath}
\usepackage{amssymb}
\usepackage{nicefrac}
\usepackage{microtype}
\usepackage{graphicx}
\usepackage{xcolor}

\newcommand{\btheta}{\boldsymbol{\theta}}

\title{Neural Bayesian Online Calibration with Deficient Simulators for MADMAX Dark Matter Experiment}

\author{
  Nabil Salama \\
  University of Hamburg \\
  Hamburg, Germany \\
  \texttt{nabil.salama@desy.de}
}

\begin{document}

\maketitle

\begin{abstract}
Scientific experiments often rely on deficient simulators and hardware whose commanded settings differ from the states realized in practice, limiting reliable optimization and inference.
A Bayesian online calibration framework is introduced for experiments with at least one directly measurable simulator output and potentially latent quantities of scientific interest.
Pre-calibration data are used to learn state-dependent effective simulator parameters, while a normalizing flow enables amortized posterior inference from measurement histories.
During calibration, measurement and positioning uncertainties are propagated, and hardware commands are selected by balancing expected improvement with information gain.
The framework is applied to the MADMAX dark matter experiment, where measured reflectivity is used to calibrate a deficient disk-booster simulator and infer the unobservable axion-signal power boost factor.
Experimental validation on the MADMAX experiment is ongoing.
\end{abstract}

\section{Introduction}

Presented in the following is an online calibration procedure for optimizing experimental configurations and inferring quantities of interest when only a deficient simulator is available and at least one of these quantities is directly measurable.
Here, a deficient simulator is one that omits physical effects which are relevant at the precision the experiment requires, so that its predictions deviate systematically from the behaviour of the real apparatus. It is this model error, rather than measurement noise or a shortage of data, that prevents conventional approaches such as numerically optimizing setup parameters or fitting them for inference from meeting the scientific requirements of the experiment.

Bayesian calibration learns simulator parameters from physical observations while representing parameter uncertainty and model discrepancy \citep{kennedy2001calibration,brynjarsdottir2014discrepancy,plumlee2017inexact}, and likelihood-free methods infer distributions over simulator parameters from real-system observations \citep{ramos2019bayessim}, including under misspecification \citep{ward2022robustnpe,wehenkel2025rope}. Deep networks have also been used to accelerate the simulation and optimization of dielectric axion haloscopes \citep{jung2022deepmadmax}. The procedure presented here combines these with amortized conditional density estimation \citep{papamakarios2016epsilonfree,greenberg2019apt,rezende2015flows} and sequential acquisition criteria from Bayesian optimization \citep{jones1998ego,hennig2012entropy,oliveira2019uncertaininputs}; its contribution is their integration into an online hardware-calibration workflow in which measurable and latent simulator outputs share effective parameters.

\section{Motivation and Background}
\label{sec:motivation}

Dark matter is an invisible form of matter whose presence is inferred from gravitational effects, while its microscopic nature remains unknown \citep{lensing-dm-paper,rotation-dm-paper}. One possible constituent is the axion, a hypothetical light particle originally associated with the Peccei--Quinn solution to the strong \textsc{cp} problem \citep{cp-theory,weinberg1978axion,wilczek1978axion,preskill1983axiondm}. The MAgnetized Disk and Mirror Axion eXperiment (MADMAX) \citep{madmax-paper} searches for axions with masses between \(40\) and \(400\,\mu\mathrm{eV}\). In a strong magnetic field, axions can convert into electromagnetic waves \citep{sikivie1983haloscope,caldwell2017dielectric}. To enhance this weak signal, MADMAX employs a booster of dielectric disks whose spacings are chosen such that the emitted waves interfere constructively \citep{caldwell2017dielectric,millar2017foundations}. The resulting signal enhancement is quantified by the power boost factor \(\beta^2\), which directly determines the experimental sensitivity \citep{millar2017foundations}. Accurate calibration of the disks and receiver antenna is therefore essential for efficiently enhancing and collecting the signal \citep{egge2020booster,madmax2024mechanical}.

\begin{figure}[ht]
    \centering
    \includegraphics[width=.6\linewidth]{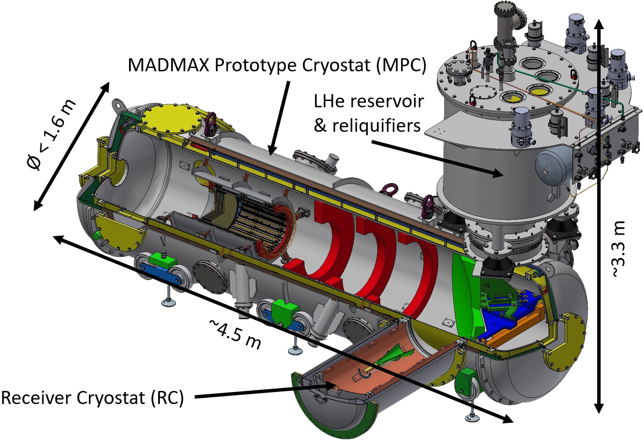}
    \caption{Illustration of the MADMAX setup with the antenna illustrated in green inside the receiver cryostat and disks illustrated in beige inside the prototype cryostat (taken from \citealp{LepplaWeber2025MADMAX}).}
    \label{fig:madmax_illustration}
\end{figure}

\label{sec:application}
For MADMAX, the power boost factor \(\beta^2=U_1\) is a hidden quantity that cannot be measured directly, whereas the reflectivity \(\Gamma=\hat U_0\) of the experimental setup is directly measurable. Initial configurations \(c_0\) can be obtained from optical and geometrical considerations for the antenna and from a deficient simulator \(S=(U_0,U_1)\) for the disk spacings. MADMAX therefore constitutes an instance of the problem class considered in this work and motivates the calibration procedure developed below. Its application to the experimental setup is ongoing and will be reported separately.

\section{Problem Setup}
\label{sec:problem}

The MADMAX experiment motivates the definition of a class of experiments and use cases for which the procedure described here might be useful. Such an experiment has moving parts that can be commanded programmatically with a command $c$ to move into a true physical state $x$, at which a quantity $\hat U_0(x)$ can be directly measured. That quantity must be simulatable, if only imperfectly, by a simulator $S$, which may also simulate further quantities $S(x,\theta)=(U_0(x,\theta), U_1(x,\theta),...)$, which may not be directly measurable.
The use case, as with many scientific experiments, is calibrating the experimental setup with regard to an experimental quantity $U_i$, which may or may not be the measurable quantity. This means aligning the hardware to optimize $U_i$ and, if $i\neq0$, also inferring it.

Usually, experimental setups are found from optical or geometrical considerations or by fitting models. If only deficient simulators are available, any setup configuration selected from this approach has not been fully optimized, and a quantity $U_i$ that is not directly measurable has not been obtained to the necessary precision either. This is mainly due to two central issues. First, optimization or parameter fitting can be biased because the best-fitting simulator parameters need not absorb the model discrepancy in a physically meaningful or identifiable way \citep{kennedy2001calibration,brynjarsdottir2014discrepancy,plumlee2017inexact}; for MADMAX, finite-size and three-dimensional electromagnetic effects, diffraction, antenna reflections and losses cause departures from idealized one-dimensional simulations \citep{knirck2019threed,egge2020booster,egge2024reciprocity}. Second, any experimental configuration obtained from calculations can only be reproduced in the laboratory up to the positioning accuracy and repeatability of the mechanical system \citep{egge2020booster,madmax2024mechanical}.

\section{Method}
\label{sec:method}

At its core, the presented method uses physical measurements to learn a calibrated effective simulator from a deficient one \citep{kennedy2001calibration,ramos2019bayessim,ward2022robustnpe}. It consists of three stages: a pre-calibration data campaign, training the machine learning models, and an iterative calibration run yielding an optimized, uncertainty-aware estimate of the experimental quantity of interest.
For a setup with $N$ calibratable hardware degrees of freedom, the simulator parameters $\theta=(\theta_0,\theta_1)$ are divided into state-describing parameters $\theta_0\in\mathbb{R}^N$, with the same dimensionality as the state $x$, and additional physics-describing parameters $\theta_1\in\mathbb{R}^M$. The calibrated effective simulator is the deficient simulator evaluated at effective, not necessarily physical, parameters $\theta(x)$ that describe the true, potentially unmeasurable, quantity $\hat U_i(x)$. These predictive calibration variables need not be identifiable with the corresponding physical parameters \citep{brynjarsdottir2014discrepancy,plumlee2017inexact}. The procedure aims to infer the effective parameter representation $\theta^*(x)$ and find the optimal setup $x^*$ with respect to an objective $J(U_i(x,\theta^*))$, such that $U_i(x^*,\theta^*(x^*))$ approximates $\hat U_i(x^*)$ with low uncertainty. The resulting $x^*$ is verified by how well $U_0(x^*,\theta^*(x^*))$ approximates the measurable quantity $\hat U_0(x^*)$.

\subsection{Pre-Calibration Data Campaign}
First, a command region $\mathcal{C}\subseteq\mathbb{R}^N$ constraining possible hardware setups is defined, and the setup is brought into an initial position $c_0\in\mathcal{C}$ by optical or geometrical considerations or by fitting the deficient simulator. Around that position, a population distribution $c_j\sim P(c)\vert_\mathcal{C}$ is defined, from which possible setup configurations $c_j$ can be drawn.

Then, a data campaign of length $J$ is recorded. For each drawn sample $c_j$, the record $D_{\mathrm{cal},j}=(c_j, x_j, \hat U_{0,j}(x_j))$ is saved. Here, $c_j$ is the commanded position, $x_j$ the physical position, and $\hat U_{0,j}(x_j)$ the measurable quantity.

Additionally, a configuration landing model is obtained from repeatedly driving the hardware to configuration $c$ and measuring $x_i$ for each approach. From there, observed configuration covariance $\Sigma_x$ as well as the mean $\mu_L$ and covariance $\Sigma_L$ of the landing residual $L=c-x$ can be calculated to obtain the landing model $p_L(x|c)=\mathcal{N}(x;c-\mu_L,\Sigma_L)$. The realized query location is therefore random even though the commanded configuration is known \citep{oliveira2019uncertaininputs}. $\Sigma_x$ is propagated through the position sensitivity of the $U_0$ measurement likelihood.

Similarly, a measurement uncertainty $\Sigma_{\hat U_0}$ on $\hat U_0(x)$ is obtained from repeated measurements at a fixed landing $x$.
From this, an observation model is defined as $u_0(x,\theta)=U_0(x,\theta) + \epsilon_U$ with likelihood
\begin{align}
    p(u_0 \mid \theta,x)
    = \mathcal{N}(U_0(x,\theta), \Sigma_{\hat U_0} + J_x\Sigma_x J_x^\top) \label{eq:measurement_likelihood}
\end{align}
where the position Jacobian $J_x=\partial U_0/\partial x \mid_x$ is used to propagate the uncertainty $\Sigma_x$. This is a first-order propagation of input covariance through the measurement model \citep{jcgm2008gum} and assumes measurement and landing errors to be independent.

\subsection{Pre-Calibration Training}

The prior distribution $p(\theta|x)$ of effective configurations conditioned on a configuration needs to be learned from the previously recorded data campaign $D_\mathrm{cal}=\left( D_{\mathrm{cal},j} \right)_{1,...,J}$.

This conditional prior $p_\psi(\theta|x)$ is defined by a prior model with parameters $\psi$. For record $j$, the measurement likelihood $p(u_{0,j}|\theta,x)$ from Equation~\ref{eq:measurement_likelihood} can be used to define a maximum likelihood objective, to which a prior model (e.g., a Gaussian mixture) can be fitted.
\begin{align}
\widehat\psi=
\operatorname*{arg\,max}_{\psi}\sum_{j=1}^{J}
\log\left[\int
p(u_{0,j}|\theta,x)
\,p_\psi(\theta\mid x_j)
\,\mathrm{d}\theta
\right]
\end{align}
From this fitted $p_{\hat\psi}(\theta^*|x)$, effective parameter samples $\theta^*$ can be drawn. From here on, the star will be dropped and $\theta$ is being understood as the fitted $x$-dependent $\theta^*$. The more general posterior over all measurements taken is then:
\begin{align}
    \pi(\theta_{1:J} \mid D_\mathrm{cal}) \propto
    \prod_{j=1}^{J} p_{\hat\psi}(\theta_j \mid x_j) p(u_{0,j} \mid \theta_j, x_j)
\end{align}
Next, a conditional normalizing flow $F_\phi$ is trained as an amortized posterior estimator, yielding $q_\phi(\theta \mid D)$ \citep{rezende2015flows,papamakarios2016epsilonfree,greenberg2019apt}, with calibration particles $\theta_i=F_\phi(z_i;h_D)$, where $z_i\sim \mathcal{N}(0,I)$ is the base sample of the flow and $h_D$ the encoded history $D$. This means that a network is trained to provide a distribution over plausible parameters for some observed measurement history.

Training data for the flow is generated synthetically by drawing from the same conditional prior, simulator, and measurement likelihood, that is later being used in the iterative calibration run. That way, a measurement history $D_i$ of length $K_i$ is generated as network input and the parameter sets $\btheta$ generating that history are retained as labels. Thus, one training example $\mathcal{T}_i=(D_i,\btheta_i, K_i)$ contains history $\{D_{ik}=(c_{ik},x_{ik},u_{0,{ik}})\}_{k=1}^{K_i}$ and parameter sets $\btheta_i=(\theta_{i1},\ldots,\theta_{i K_i})$ with $\theta_{ik}\in \mathbb{R}^{N+M}$.
The training data is obtained from the available distributions:
\begin{align}
    c \sim P(c), \qquad x \sim p_L(\cdot \mid c), \qquad \theta \sim p_\psi(\cdot \mid x_i) \qquad u_0 \sim p(\cdot \mid \theta,x)
\end{align}
These synthetic histories approximate the data-generating process encountered during calibration. Their adequacy must nevertheless be validated against held-out real measurements, because neural posterior estimators can become biased or overconfident when the simulator is misspecified \citep{brynjarsdottir2014discrepancy,ward2022robustnpe,wehenkel2025rope}.
The network structure consists of an encoder that transforms the variable-length history $D$ into a context vector $h_D$ and a normalizing flow that is used to propose parameter sets from the posterior distribution.

For each record $D_{ik}$, encoder inputs are (i) the nominal simulated measurement $U_0(x_{ik},\theta_{ik})$, (ii) the residual $r_{ik}=u_{0,ik}-U_0(\theta_a,x)$ for an arbitrary global anchor $\theta_a$, (iii) a tangent matrix $A_{ik}=\partial U_0(\theta_a,x)/\partial \eta \mid_{\eta=T(\eta_a)}$ for a transformation $T$ mapping into unconstrained normalized space, (iv) and optionally any other suitable experimental quantity $Q_{ik}$. This input set describes for position $x_{ik}$, how the effective simulator behaves, how it differs from a measurement, and how parameters change around the anchor region.
For multidimensional or semantically ordered quantities $U_0$ (e.g. spectra), a convolutional neural network may be used as the first encoder stage to produce an embedding $e_{ik}$. The history of training example $i$ up to record $k$ is described by $e_{ik}=\sum_{n=1}^k e_{in}/K_i$. This sums the embeddings only up to step $k$ for a total history length $K_i$ and is thus not the mean, but a quantity that grows with the amount of context information. This is a Deep-Sets-style permutation-invariant aggregation \citep{zaheer2017deepsets} and is appropriate only where the ordering of measurements carries no information. This vector, together with $k$ or any other further information, is passed to a context network, which forms the second encoder stage and produces a fixed-length context vector $h_{D_i}$ that serves as the input for the normalizing flow. The flow evaluates the posterior density $q_\phi(\theta_{ik} \mid e_{ik}, h_{D_i}, x_{ik})$ at training with batch size $B$ and label $\theta_{ik}$ for a history up to length $K_i$, which is used with the negative log-likelihood loss as the objective:
\begin{align}
\mathcal{L}(\phi)
=-\frac{1}{B}\sum_{i=1}^{B}
\log q_{\phi}(T(\boldsymbol{\theta}_i) \mid D_i),
\qquad
q_{\phi}
(\boldsymbol{\eta}_i\mid D_i)
=\prod_{k=1}^{K_i}
q_\phi(\eta_{ik} \mid e_{ik},h_{D_i},x_{ik}).
\end{align}
\subsection{Calibration Run}
In the calibration run, the previously learned effective prior $p_\psi(\theta \mid x)$, the learned posterior $q_\phi(\theta \mid D)$, and the measurement model are being used to iteratively find a configuration $x^*$ that optimizes the quantity $U_i$. The procedure exploits the entire calibration measurement history $D$ of length $J$ to achieve that.
That history is initialized with measurements at targets $\{c_0, c_0\pm h_le_l \mid l=1,\dots,N\}$ for a small step size $h_l$ and component basis vector $e_l$. For every measurement, $D_j=(c_j,x_j,\hat U_0(x_j))$ is recorded.

To obtain the quantity $U_i(x_j)$ and its posterior distribution at a measurement position $x_j$, draw $K$ parameter samples $\Theta_j=\theta_{j,1:K}\sim q_\phi (\cdot\mid D)$ from the normalizing flow. The particles are corrected with the learned conditional prior and measurement likelihood by $p(\Theta_j \mid x_{1:j})p(D\mid \Theta_j,x_{1:j})/q_\phi(\Theta_j\mid D)$, or in log weights for numerical stability:
\begin{align}
a_k=
\sum_{j=1}^{J}
[\log p(\theta_{j,k}\mid x_j)
+\log p(u_{0,j}\mid\theta_{j,k},x_j)]
-\log q_{\phi}(\theta_{jk}\mid D).
\end{align}
These are then normalized to $w_k=\exp(a_k-a_m)/\sum_i\exp(a_i-a_m)$ with maximum $a_m=\max_i a_i$.

The posterior distribution of $U_i(\theta,x_j)$ is then represented by the weighted collection $\{(U_i(\theta_k,x_j),w_k)\}_{k=1}^{K}$ from which the final value at position $x_j$ is calculated as $U_i(x_j)=\sum_k w_k U_i(\theta_k,x_j)$.

To select the next setup target to append to the history, for each potential next command $c^+$, a utility improvement and information gain are calculated. Based on these two factors, the next physical target is selected, and the hardware is driven to that new state.
First, one of the experimental quantities $U_i$ has to be chosen, and a utility $J(U_i)$ needs to be defined on it, which is maximal for the best value of $U_i$.
For a possible next command, an expected-improvement criterion \citep{jones1998ego} is used, $EI(c^+)=\sum_k w_k[\bar J_k(c^+) - \bar J_k(c^*)-\epsilon_J]_+$, where $c^*=\arg \max_j J(U_i(x_j))$ is the best command in the calibration history, $[a]_+=\max(a,0)$, and $\epsilon_J$ a suitable improvement scale. The expected utility over $S$ landing scenarios is $\bar J_k(c)=\mathbb{E}_{x^+\sim p_L(\cdot\mid c)}[J(U_i(\theta_k,x^+))]$, which accounts for uncertainty in the realized physical query location \citep{oliveira2019uncertaininputs}.

For the information gain, let $\mathcal{C}_\mathrm{cand}\subset\mathcal{C}$ be a finite set of candidate commands. For each candidate command $c^+\in\mathcal{C}_\mathrm{cand}$, proposed particle $\theta_k$, and landing scenario $s=1,\ldots,S$, draw a state $x_{kcs}^{+}\sim p_L(\cdot\mid c)$ and effective parameters $\theta_{kcs}^{+}\sim p_{\hat\psi}(\cdot\mid x_{kcs}^{+})$, and define $J_{kcs}=J(U_i(\theta_{kcs}^{+},x_{kcs}^{+}))$ and $Z_{ks}=\operatorname*{arg\,max}_{c\in\mathcal{C}_\mathrm{cand}}J_{kcs}$. The current probability that command $c$ is optimal is $p_c=\sum_k w_k S^{-1}\sum_s\mathbf{1}_{Z_{ks}=c}$, with entropy $H_0=-\sum_c p_c\log p_c$. For a possible next command $c^+$, draw a hypothetical measurement from $p(u_0^+\mid D,c^+)=\sum_{k,s}(w_k/S)p(u_0^+\mid\theta_{k,c^+,s}^{+},x_{k,c^+,s}^{+})$ and reweight the candidate worlds by $\widetilde w_{ks}\propto(w_k/S)p(u_0^+\mid\theta_{k,c^+,s}^{+},x_{k,c^+,s}^{+})$. If $H(u_0^+,c^+)$ denotes the resulting entropy, the expected information gain is
\begin{align}
I(c^+)=H_0-\mathbb{E}_{p(u_0^+\mid D,c^+)}
\left[H(u_0^+,c^+)\right],
\end{align}
which favors measurements expected to resolve which command maximizes $J$, following the objective of learning the location of the optimizer rather than only seeking a high immediate objective value \citep{hennig2012entropy}.

The expected improvement and information gain are combined into one acquisition score. A variable $\alpha_t\in[0,1]$ is defined to control the balance between improving $U_i$ and reducing uncertainty about the optimal command. The coefficient $\alpha_t$ may be decreased from one to zero to move from improvement-focused to information-focused selection. The next command $c_{t+1}$ is thus selected as:
\begin{align}
c_{t+1}&=
\operatorname*{arg\,max}_{c^+\in\mathcal{C}_\mathrm{cand}}
\mathcal{A}_t(c^+),&
\mathcal{A}_t(c^+)
&=\alpha_t \frac{EI(c^+)}{EI(c^+)+\epsilon_J}
+(1-\alpha_t)\frac{I(c^+)}{\log|\mathcal{C}_\mathrm{cand}|}.
\end{align}
Only commands whose landing distribution $p_L(x\mid c^+)$ remains inside the region covered by the learned prior $p_{\hat\psi}(\theta\mid x)$ should be considered.

\section{Assumptions and Limitations}
\label{sec:assumptions}
If the quantity $U_i$ that should be optimized differs from the directly measurable quantity $U_0$, then the method applies only to situations in which a simulator is available for which the same parameter set $\hat\theta$ describes both $U_0(\hat\theta,x)$ and $U_i(\hat\theta,x)$.

It is further assumed that every measurement $\hat U_0(x)$ lies in the codomain of the simulator $U_0(\theta,x)$, and similarly for $U_i$. This simulator-specific support-overlap assumption is stronger than approaches that introduce an explicit observation-space discrepancy or learn a calibration map between simulated and real observations \citep{brynjarsdottir2014discrepancy,ward2022robustnpe,wehenkel2025rope}.

\bibliographystyle{plainnat}
\bibliography{references}

\end{document}